# Field models and numerical dosimetry inside an extremely-low-frequency electromagnetic bioreactor: the theoretical link between the electromagnetically induced mechanical forces and the biological mechanisms of the cell tensegrity

Maria Evelina Mognaschi[1], Paolo Di Barba[1], Giovanni Magenes[1,2], Andrea Lenzi[3], Fabio Naro[4] and Lorenzo Fassina[1,2*]

**Abstract**

We have implemented field models and performed a detailed numerical dosimetry inside our extremely-low-frequency electromagnetic bioreactor which has been successfully used in *in vitro* Biotechnology and Tissue Engineering researches. The numerical dosimetry permitted to map the magnetic induction field (maximum module equal to about 3.3 mT) and to discuss its biological effects in terms of induced electric currents and induced mechanical forces (compression and traction). So, in the frame of the tensegrity-mechanotransduction theory of Ingber, the study of these electromagnetically induced mechanical forces could be, in our opinion, a powerful tool to understand some effects of the electromagnetic stimulation whose mechanisms remain still elusive.

**Keywords:** Electromagnetic field models; Numerical electromagnetic dosimetry; Magnetic induction field; Induced electric field; Induced electric currents; Induced mechanical forces; Tensegrity

## Introduction

The research about the biological effects caused by electromagnetic fields (EMFs) has been of great interest in the past decades. In particular, extremely-low-frequency EMFs (ELF-EMFs), with frequency up to 300 Hz and continuously irradiated by civil and industrial appliances, have been investigated to clarify their possible biological effects on the population unceasingly exposed to them; to this regard, epidemiological studies have shown a relation between the environmental ELF-EMFs and the onset of leukemia (tumor of the lymphoid tissue) (Kheifets et al., 2010) or Alzheimer's disease (neurodegenerative disorder in the non-lymphoid brain tissue) (Davanipour et al., 2007; Huss et al., 2009; Maes and Verschaeve, 2012).

In both lymphoid and non-lymphoid tissues, the cells regulate the flow of ionic currents across their plasma membrane and internal membranes through specific ion channels, so that one of the simplest ways to affect a biological system is to induce a change in its ionic fluxes (for instance, via an ELF-EMF exposition that elicits conformational changes in the ion channel proteins and modifies, in particular, the calcium currents and the cytosolic calcium concentration), as it is well known that an increased calcium flux can trigger numerous biochemical pathways (Bawin et al., 1978; Walleczek, 1992; Balcavage et al., 1996; Pavalko et al., 2003). As a matter of fact, ELF-EMFs lead to a mitogenic effect in lymphocytes because they can modify the calcium influx (Balcavage et al., 1996; Murabayashi et al., 2004), whereas, in a non-lymphoid tissue such as the brain neuronal tissue, the proposed mechanisms about the electromagnetic stimulation are more complex and involving both ionic fluxes and alterations in the distribution and in the

\* Correspondence: lorenzo.fassina@unipv.it
[1]Dipartimento di Ingegneria Industriale e dell'Informazione, Università di Pavia, Via Ferrata 1, Pavia 27100, Italy
[2]Centro di Ingegneria Tissutale (C.I.T.), Università di Pavia, Pavia, Italy
Full list of author information is available at the end of the article functionality of membrane receptors (e.g. serotonin, dopamine, and adenosine receptors).

In particular, the ELF-EMFs decrease the affinity of the G-protein-coupled 5-HT$_{1B}$ serotonergic receptor with a consequent decreased signal transduction (Massot et al., 2000; Espinosa et al., 2006), decrease the affinity of the G-protein-coupled 5-HT$_{2A}$ serotonergic receptor (Janac et al., 2009), reduce the reactivity of the central dopamine D$_1$ receptor (Sieron et al., 2001), and increase the density of the A$_{2A}$ adenosine receptor (Varani et al., 2011) revealing, as a consequence, a possible treatment of the inflammatory trait in Alzheimer's disease via a better use of the endogenous adenosine, which is an effective brain anti-inflammatory agent when combined with its A$_{2A}$ receptor (Rosi et al., 2003; Tuppo and Arias, 2005). In addition, the adenosine receptors appear to play an important role during the *in vitro* ELF-EMF stimulation of other cell types such as neutrophils (Varani et al., 2002; Varani et al., 2003), chondrocytes and fibroblast-like synoviocytes (Varani et al., 2008; De Mattei et al., 2009).

The preceding positive biological effects, which can be described as an electromagnetic modulation of the cellular and tissue functions, have been obtained at extremely low frequencies and very low magnetic fields. In our *in vitro* experience, in order to enhance the biological effects, we have utilized a similar electromagnetic wave with a frequency of about 75 Hz (instead of the 50 Hz or 60 Hz of the electric devices), a module of the magnetic field equal to circa 3 mT (i.e. about 60-fold the intensity of the Earth magnetic field), and with a solenoids' spatial configuration to assure, where the cells are seeded, the maximum homogeneity of the magnetic field.

In particular, we have showed that an ELF-EMF stimulus could elicit a cytoprotective response in human neurons in terms of production of the neurotrophic factor sAPPalpha, promotion of the non-amyloidogenic pathways, and protection against cellular stress and oxidation (Osera et al., 2011) via enhanced expressions of the chaperone heat shock protein HSP70 and the free radical scavenger SOD-1, respectively. On the other hand, we have used the same electromagnetic bioreactor to perform bone tissue engineering experiments: to enhance the *in vitro* culture of biomaterial scaffolds, the electromagnetic stimulus was applied to increase the cell proliferation and the synthesis of type-I collagen, decorin, osteocalcin, and osteopontin, which are fundamental constituents of the physiological bone matrix (Fassina et al., 2006; Fassina et al., 2007; Fassina et al., 2008; Fassina et al., 2009; Fassina et al., 2010; Ceccarelli et al., 2013).

As a consequence, the aim of the present work is to accomplish a detailed numerical dosimetry inside our electromagnetic bioreactor in order to show the specific and effective physical stimulus transduced by the cells *in vitro*, not only by describing the local time-dependent magnetic field, but also by discussing the local hydrostatic forces (perpendicular to the cell membranes) and the local shear forces (parallel to the cell membranes), both caused by the magnetic field; in other words, we aim to frame this kind of stimulation not only under an electromagnetic viewpoint, but also under the tensegrity-mechanotransduction theory of Ingber (Mammoto and Ingber, 2010).

## Materials and methods
### Experimental setup of the electromagnetic bioreactor

The experimental setup of our electromagnetic bioreactor is based on two solenoids (i.e. air-cored Helmholtz coils) connected in series and powered by a pulse generator (Biostim SPT Pulse Generator from Igea, Carpi, Italy) (Figure 1). The solenoids have a quasi-rectangular shape (length, 17 cm; width, 11.5 cm) and their planes are parallel with a distance of 10 cm, so that the cell cultures can be placed 5 cm away from each solenoid plane. The preceding setup is based on the theory of the Helmholtz coils, that is, in order to optimize the spatial homogeneity of the magnetic field, especially in the central region where the cells are stimulated, the two coils should be supplied by the same current (i.e. with same magnitude and direction) and their dimensions and distance should be comparable (in particular, the coils' diameter and distance should be equal if the coils have a circular shape).

### Electric measurements

In order to create a finite element model, some electric measurements were performed. The coils are powered via a Burndy connector, of which two terminals are used for delivering current to the coils. Current and voltage measurements were simultaneously performed as shown in Figure 2.

The pulse generator fed the two 1000-turns coils in series by a nearly square-wave voltage (frequency equal to 75 Hz), whereas the resulting current in the coils' wire ranged from 0 to about 319 mA in 1.36 ms (under a finite element viewpoint, this current was equivalent to 0–319 A in 1.36 ms flowing in each winding) (Figure 2).

The preceding measurements were then used to estimate the resistance R and the inductance L of the coils via a custom-made script in Matlab language (The MathWorks, Inc., Natick, MA). In particular, given an applied voltage in the lumped-element RL series circuit, the script identified R and L in a current transient by minimizing an error functional based on the measured current and the estimated one (minimization via the simplex method; error functional tolerance less than $10^{-4}$; measurement error of about 2–3%). The estimated coils' parameters were R = 545 Ω and L = 595 mH. After that, in order to validate the estimated resistance R, a measurement with a digital multimeter was also carried out: R resulted equal to 548 Ω, that is, in good agreement with the estimated value. As a consequence, because of the coils were connected in series, each coil was approximately characterized by L = 298 mH and R = 272 Ω.

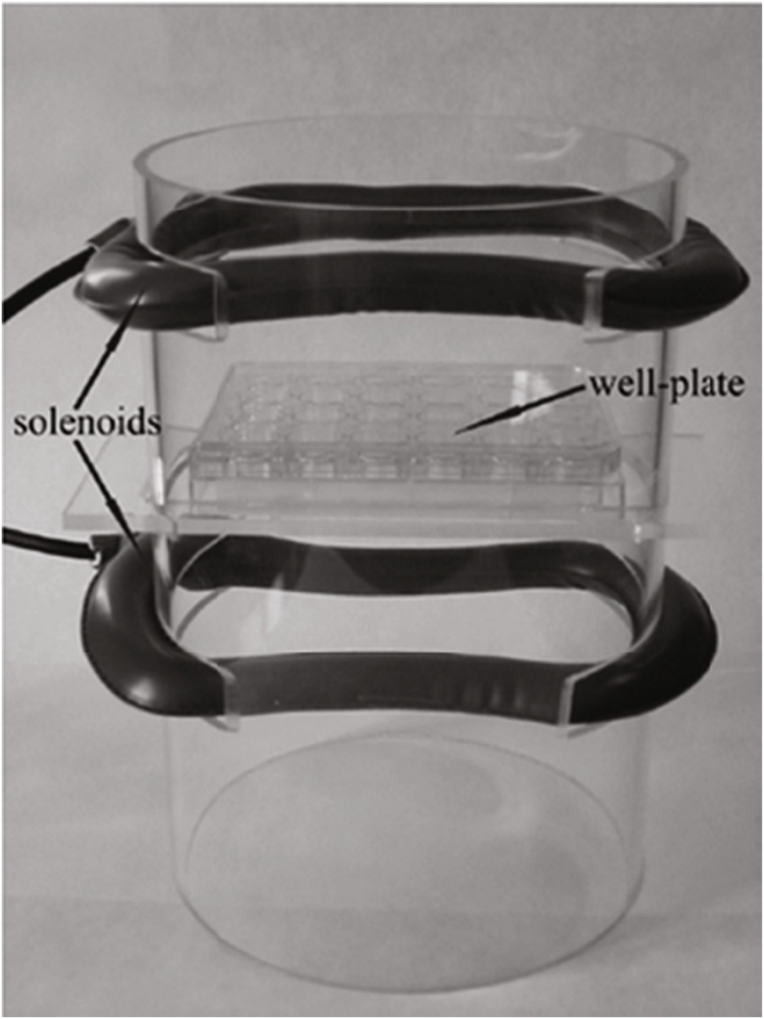

Figure 1 **Electromagnetic bioreactor.** Solenoids of the electromagnetic bioreactor with a culture well-plate in the central region.

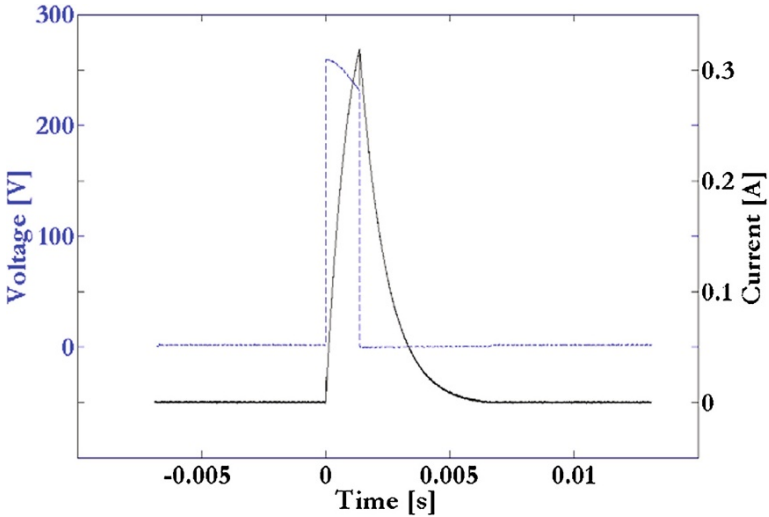

Figure 2 **Electric measurements.** Measurements of current (black continuous line) and voltage (blue dashed line).

### Finite element models

In order to simulate the magnetic field produced by the electromagnetic bioreactor, two 3D finite element models were implemented: a linear/static (Problem 1) and a linear/time-dependent (Problem 2). A third problem (Problem 3) was solved to calculate the field effects due to the metallic plates of the incubator where the bioreactor was placed during the *in vitro* experiments. The third problem was time-dependent and both linear and non-linear materials were considered.

Thanks to field symmetries, it was possible to model only 1/8 of the entire device (Figure 3), in other words, to simulate the full problem with two coils, it was sufficient to set up 1/4 of a coil and to impose specific boundary conditions (see below the Equations 3 and 5).

### Formulation of the models in terms of dual potentials

The $\bar{T}-\Omega$ method is based on a pair of dual potentials: (i) the electric vector potential $\bar{T}$ such that $\bar{\nabla} \times \bar{T} = \bar{J}$ where $\bar{J}$ is the current density and (ii) the magnetic scalar potential $\Omega$ such that $\bar{H} = -\bar{\nabla}\Omega + \bar{T}$ where $\bar{H}$ is the magnetic field. Accordingly, the magnetic problem is formulated as follows:

$$\bar{\nabla}^2 \bar{T} - \mu\sigma \frac{\partial \bar{T}}{\partial t} = -\bar{\nabla} \times \bar{J} \quad (1)$$

$$\nabla^2 \Omega - \mu\sigma \frac{\partial \Omega}{\partial t} = 0 \quad (2)$$

where σ is the electrical conductivity and μ is the magnetic permeability. The boundary conditions are (Figure 3):

$$\bar{n} \cdot \bar{T} = 0 \text{ along the planes } x = 0 \text{ and } y = 0 \quad (3)$$

$$\bar{n} \times \bar{T} = 0 \text{ elsewhere} \quad (4)$$

$$\frac{\partial \Omega}{\partial n} = 0 \text{ along the planes } x = 0 \text{ and } y = 0 \quad (5)$$

$$\Omega = const \text{ elsewhere} \quad (6)$$

If a model subregion contains a ferromagnetic material, the relevant magnetic permeability depends on the unknown field and, for this reason, the Equations 1 and 2 are non-linear and can be solved by an iterative procedure. The $\bar{T}-\Omega$ method is cost-effective in the case of 3D models because the vector potential is defined only in the conductive subregions, while the scalar potential is defined elsewhere. As a consequence, using a numerical grid to discretise the field domain, there are three unknowns per node in the conductive subregions, whereas one unknown in the other nodes.

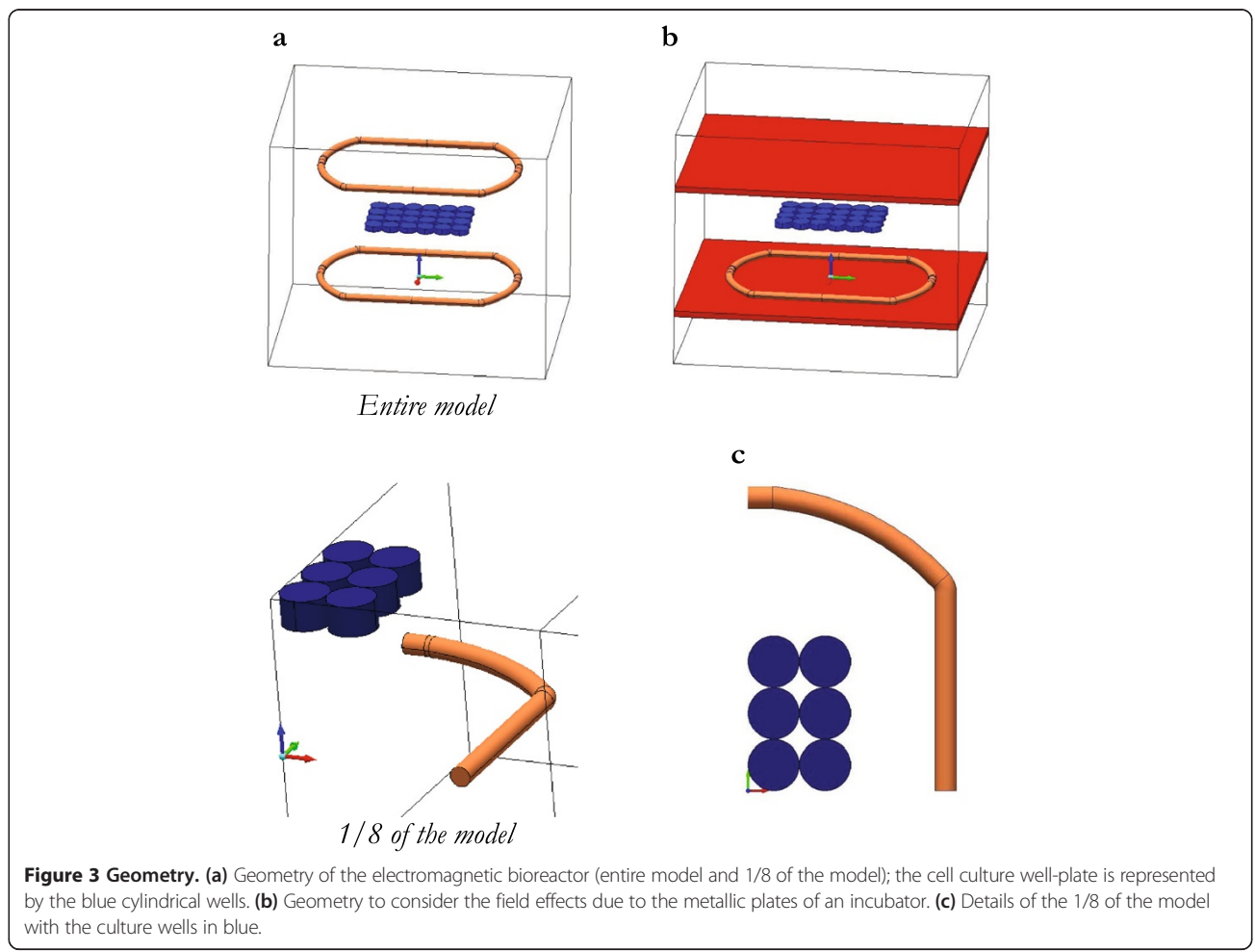

**Figure 3 Geometry.** (a) Geometry of the electromagnetic bioreactor (entire model and 1/8 of the model); the cell culture well-plate is represented by the blue cylindrical wells. (b) Geometry to consider the field effects due to the metallic plates of an incubator. (c) Details of the 1/8 of the model with the culture wells in blue.

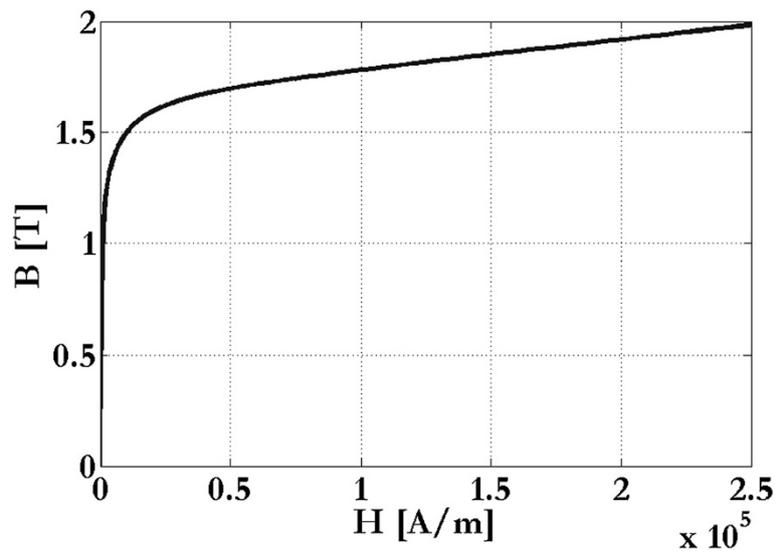

**Figure 4 Metallic plates.** B-H curve of the magnetic martensitic stainless steel considered.

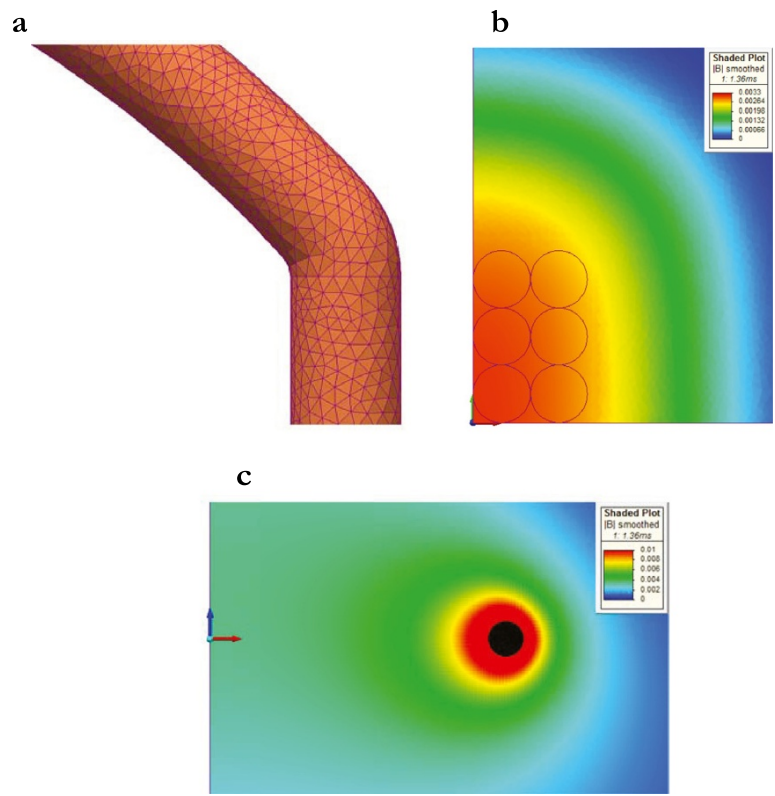

**Figure 5 Mesh and magnetic induction. (a)** Detail of the coil mesh. **(b)** Module of the magnetic induction in the plane z = 5 cm, that is, in the central region of the electromagnetic bioreactor where the cell cultures were stimulated, and for t = 1.36 ms when the coil current was maximum (the cell cultures were placed inside wells here represented by thin black circles). In this region, the cells appeared homogeneously irradiated. **(c)** Module of the magnetic induction in the plane y = 0 and for t = 1.36 ms when the coil current was maximum. The coil is represented in black.

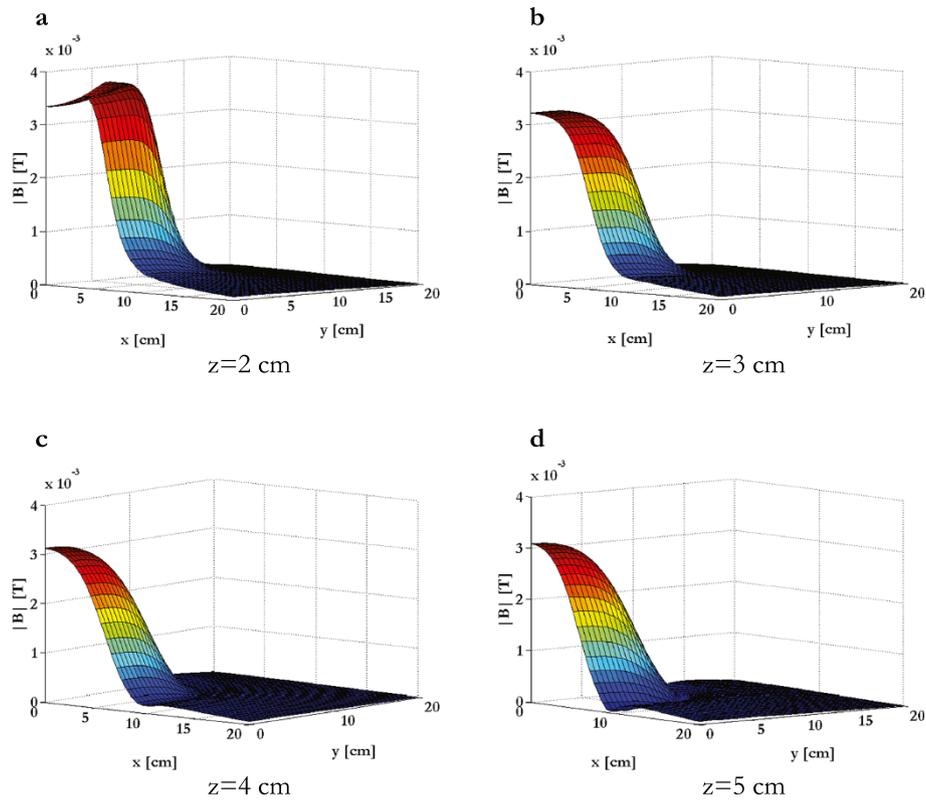

**Figure 6 Magnetic induction.** Module of the magnetic induction for t = 1.36 ms and evaluated in parallel planes: **(a)** z = 2 cm, **(b)** z = 3 cm, **(c)** z = 4 cm, **(d)** z = 5 cm.

We have solved the Problems 1, 2, and 3 by the $\bar{T}-\Omega$ method implemented in the finite element tool MagNet (version 7, Infolytica Corporation, Montréal, Canada) and running on a 64 bit PC with 8-core CPU and 16 GB of RAM.

### Static problem without culture medium (Problem 1)
A 3D linear/static problem was implemented to assess the geometry of the electromagnetic bioreactor and its electric properties. For this reason, it was not necessary to include a culture well-plate between the coils. In the static approximation, the time derivatives are null and the Equations 1 and 2 become:

$$\bar{\nabla}^2 \bar{T} = -\bar{\nabla} \times \bar{J} \qquad (7)$$

$$\nabla^2 \Omega = 0 \qquad (8)$$

The total current flowing in the 1/4 coil was assumed equal to the measured peak current of 319 A (as discussed above).

### Time-dependent problem with culture medium (Problem 2)
A 3D linear/time-dependent problem was implemented to calculate the magnetic field in the real electromagnetic bioreactor. The current flowing in the 1/4 coil was considered time-dependent as shown in Figure 2. A culture well-plate with real dimensions was included and filled by a physiological saline solution with an electrical conductivity of 1.84 S/m (Figure 3). The problem was solved according to the Equations 1, 2, 3, 4, 5, and 6.

**Table 1 $B_z$ and $B_x$ for y = 0 cm, z = 4.5 cm, and t = 1.36 ms ($B_y$ was negligible)**

|  | x = 0 cm | x = 3 cm | $|\Delta B| = |B_i(x=0\ cm) - B_i(x=3\ cm)|$ [mT] | $|\Delta B|/B_i$ (x = 0 cm) [%] |
|---|---|---|---|---|
| $B_z$ [mT] | 3.1 | 2.73 | 0.37 | 11.9% |
| $B_x$ [mT] | Negligible | 0.14 | 0.14 | ——— |

**Table 2 $B_z$ and $B_x$ for y = 4.5 cm, z = 4.5 cm, and t = 1.36 ms ($B_y$ was negligible)**

|  | x = 0 cm | x = 3 cm | $|\Delta B| = |B_i(x=0\ cm) - B_i(x=3\ cm)|$ [mT] | $|\Delta B|/B_i$ (x = 0 cm) [%] |
|---|---|---|---|---|
| $B_z$ [mT] | 2.86 | 2.44 | 0.42 | 14.6% |
| $B_x$ [mT] | Negligible | 0.14 | 0.14 | ——— |

### Effects of metallic plates near the electromagnetic bioreactor (Problem 3)

In order to calculate the field effects due to the metallic plates of an incubator, a 3D time-dependent problem was solved according to the Equations 1, 2, 3, 4, 5, and 6 and with a time-stepping procedure to iteratively correct the magnetic permeability (Figure 3b). Because of the plates can be made by different materials, two standard alloys were then considered: (i) a non-magnetic austenitic stainless steel with electrical conductivity $\sigma = 1.3 \times 10^6$ S/m and (ii) a magnetic martensitic stainless steel with electrical conductivity $\sigma = 1.3 \times 10^6$ S/m and with a B-H curve as shown in Figure 4.

## Results
### Problem 1

The finite element mesh consisted of about $5.2 \times 10^6$ tetrahedrons and the relative field solution is shown in Figure 5 for t = 1.36 ms when the coil current was maximum (we have adopted an orthogonal Cartesian reference system with the x, y, and z axes in red, green, and blue, respectively). The maximum module of the magnetic induction was almost homogeneously equal to about 3.3 mT in the central region of the electromagnetic bioreactor (plane z = 5 cm) where our Tissue Engineering cultures were centered and stimulated (Fassina et al., 2006; Fassina et al., 2007; Fassina et al., 2008; Fassina et al., 2009; Fassina et al., 2010; Saino et al., 2011; Osera et al., 2011; Ceccarelli et al., 2013). So, we could affirm that, in this region, the *in vitro* cell cultures appeared subjected to an almost homogeneous field.

In order to assess our finite element implementation with an internal control, the inductance L and the resistance R of the coils were calculated and compared with the measured ones: L resulted equal to 369 mH in agreement with its measure (L is a function of the magnetic energy $E_m$; $E_m = 4.69 \times 10^{-3}$ J in the present model) (Stratton, 1941; Panofsky and Phillips, 1962), whereas R was equal to 278 Ω in very good concordance with the measured value (R is a function of the Joule losses P; P = 7.06 W in the present model) (Stratton, 1941; Panofsky and Phillips, 1962).

In Figure 6 the module of the magnetic induction was evaluated in parallel planes (z equal to 2, 3, 4, and 5 cm) for t = 1.36 ms when the coil current was maximum. In the central region of the electromagnetic bioreactor (plane z = 5 cm), where the cells were centered and stimulated, the magnetic induction was also mainly parallel to the z axis, that is, its $B_z$ component was predominant and the $B_x$, $B_y$ components were negligible ($|B_x|$ and $|B_y|$ less than $10^{-5}$ T).

Table 3 $B_z$ and $B_y$ for x = 0 cm, z = 4.5 cm, and t = 1.36 ms ($B_x$ was negligible)

|  | y = 0 cm | y = 4.5 cm | $|\Delta B| = |B_i(y = 0$ cm$) - B_i(y = 4.5$ cm$)|$ [mT] | $|\Delta B|/B_i$ (y = 0 cm) [%] |
|---|---|---|---|---|
| $B_z$ [mT] | 3.1 | 2.86 | 0.24 | 7.7% |
| $B_y$ [mT] | Negligible | 0.07 | 0.07 | —— |

During Biotechnology and Tissue Engineering experiments, when the presence of biomaterials could be fundamental, it is of importance to consider the thickness of the culture scaffold. As a consequence, we have also calculated the magnetic induction in the plane z = 4.5 cm in order to define and characterize a useful thickness around the central region of the electromagnetic bioreactor (the symmetric plane z = 5.5 cm showed the same results due to the model symmetries). In particular, the magnetic induction was calculated considering the real number and dimensions of standard culture wells (matrix of 3 × 2 wells in the 1/8 of the model; well diameter equal to 1.5 cm) (Figure 3); the results about the magnetic induction and its variations among the culture wells are reported in the Tables 1, 2, 3 and 4 for t = 1.36 ms when the coil current was maximum.

### Problem 2

We have implemented the same finite element mesh as in Problem 1 in order to study a culture well-plate with real dimensions and filled by a physiological saline solution (electrical conductivity of 1.84 S/m). The temporal pattern of the magnetic induction module and its time derivative (which is related to the induced voltage and to the induced electric field) were evaluated at the center of a culture well (Figure 7). These results were in very good agreement with the measures showed in our preceding Tissue Engineering works involving the present electromagnetic bioreactor (Fassina et al., 2006; Fassina et al., 2009).

### Problem 3

To calculate the field effects due to the metallic plates of an incubator, we have considered both austenitic and martensitic steel alloys which are non-magnetic and magnetic, respectively. As shown in Figures 8 and 9, at the center of a culture well, the austenitic plates did not affect the temporal pattern of the magnetic induction module, whereas the martensitic ones doubled it.

Table 4 $B_z$ and $B_y$ for x = 3 cm, z = 4.5 cm, and t = 1.36 ms ($B_x$ was negligible)

|  | y = 0 cm | y = 4.5 cm | $|\Delta B| = |B_i(y = 0$ cm$) - B_i(y = 4.5$ cm$)|$ [mT] | $|\Delta B|/B_i$ (y = 0 cm) [%] |
|---|---|---|---|---|
| $B_z$ [mT] | 2.72 | 2.44 | 0.28 | 10.3% |
| $B_y$ [mT] | Negligible | 0.095 | 0.095 | —— |

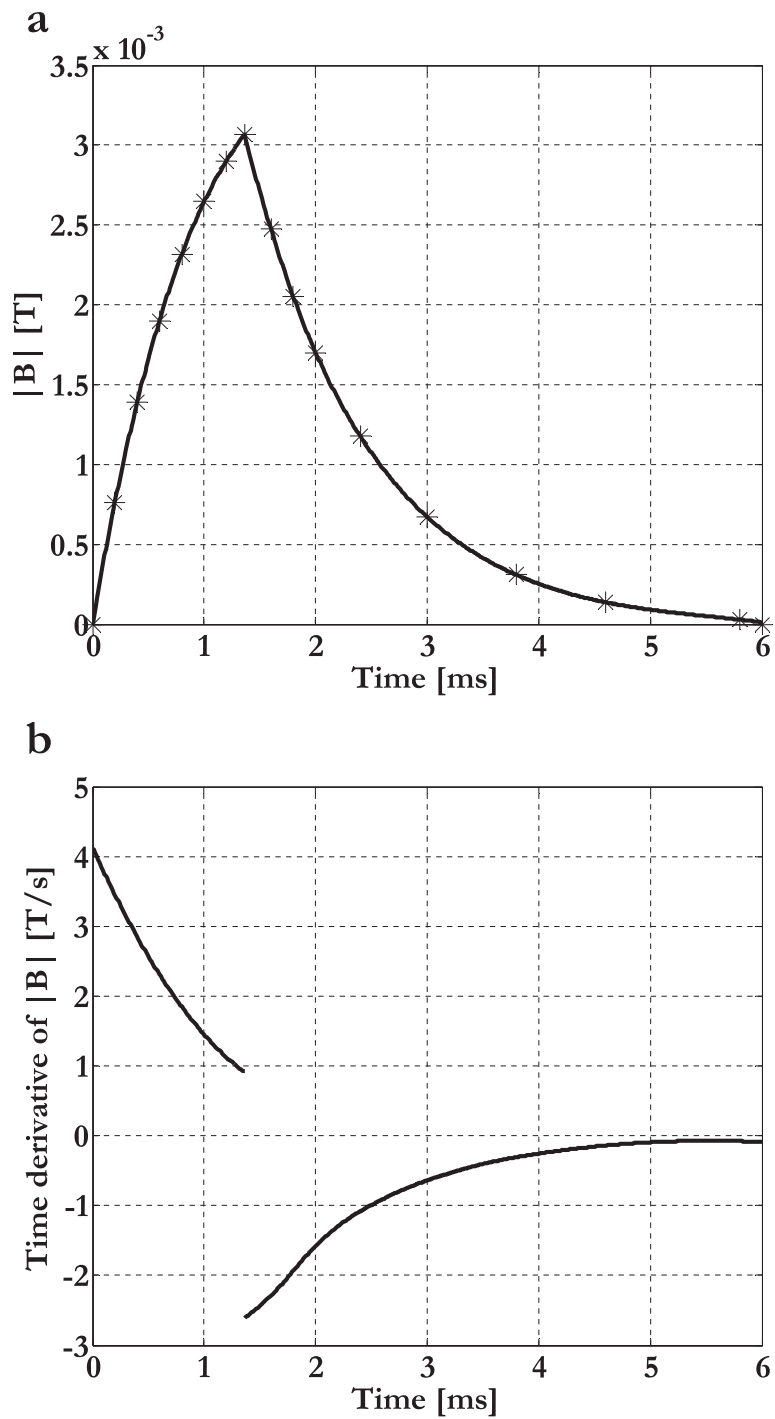

**Figure 7 Temporal pattern. (a)** Temporal pattern of the magnetic induction module at the center of a culture well for z = 4.5 cm (simulated data in asterisks with interpolation). **(b)** Time derivative of the magnetic induction module.

### Induced electric currents and induced mechanical forces inside the culture wells

According to the Faraday-Neumann-Lenz and Lorentz laws (Feynman et al., 1964), inside the cylindrical culture wells, the time varying and homogeneous magnetic induction (frequency = 75 Hz) generated a concentric and planar distribution of induced electric currents with corresponding induced distribution of radial mechanical forces: in the temporal range 0–1.36 ms the magnetic induction was arising, the currents clockwise, and the radial mechanical forces inwardly directed (compression), whereas, on the contrary, during the temporal range 1.36–6 ms, the magnetic induction was decreasing, the currents anticlockwise, and the radial mechanical forces outwardly directed (traction) (Figures 7 and 10). In particular, in order to evaluate the maximum compression and the maximum traction, we have analytically calculated the induced current density and the induced force in t = 0.64 ms and in the neighborhood of t = 1.36 ms when the coil current was maximum (Figures 7 and 10c, Table 5).

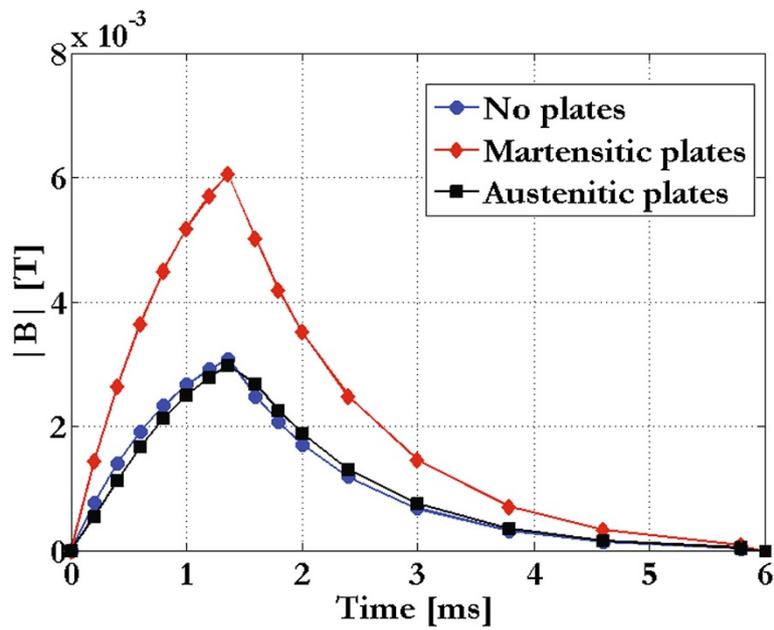

Figure 8 **Temporal pattern.** Temporal pattern of the magnetic induction module inside an incubator with austenitic or martensitic plates.

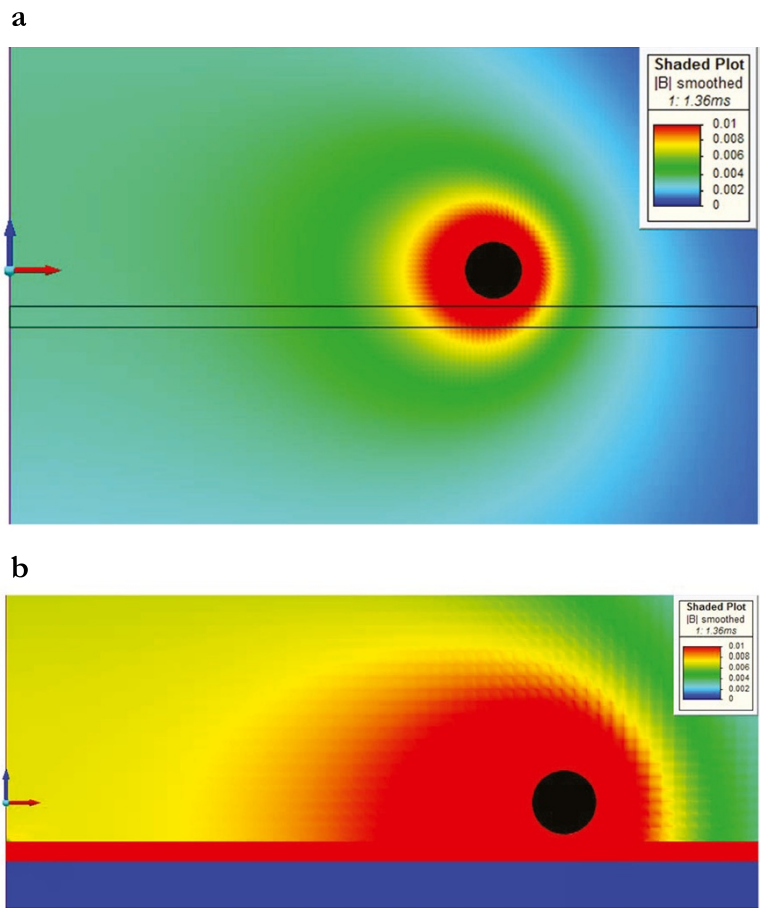

Figure 9 **Magnetic induction.** Module of the magnetic induction (plane y = 0; t = 1.36 ms) for austenitic **(a)** and martensitic **(b)** steel plates. The coil is represented in black.

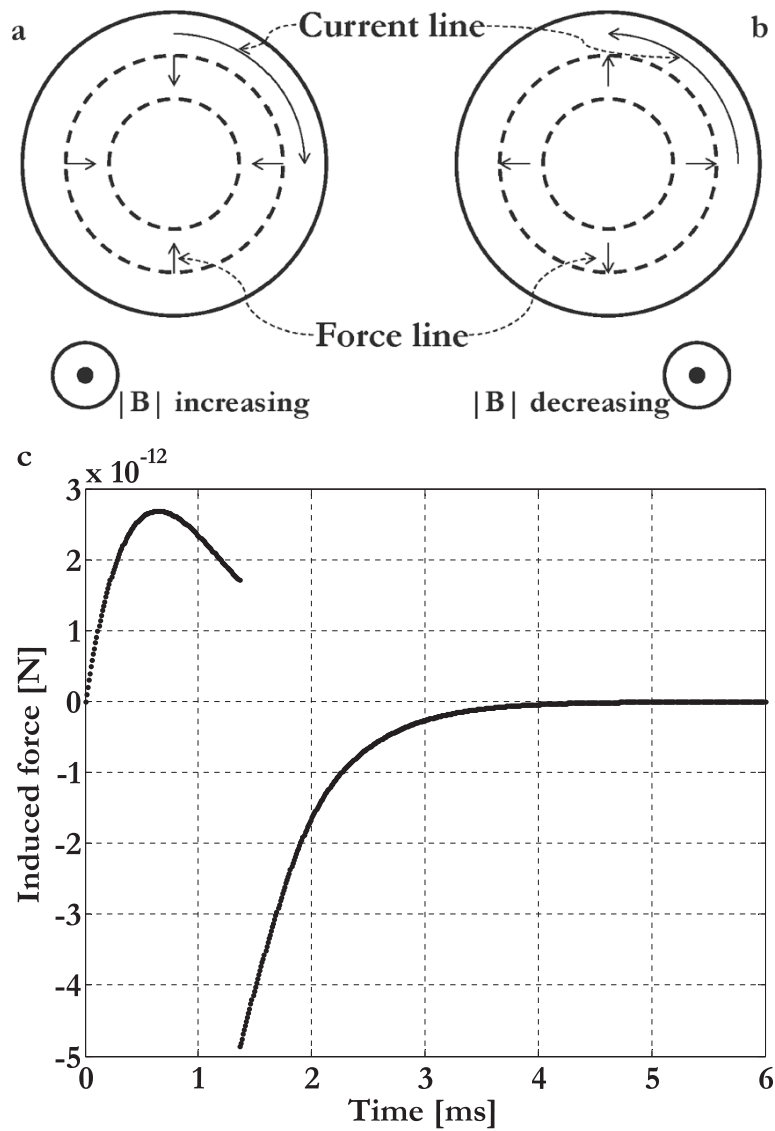

**Figure 10 Induced electric currents and induced mechanical forces.** Induced electric currents (the actual current direction is shown) and induced mechanical forces inside the culture wells during the temporal ranges 0–1.36 ms **(a)** and 1.36–6 ms **(b)**. Temporal pattern of the induced force inside the culture wells during the time range 0–6 ms (sign convention: compression force > 0 N, traction force < 0 N) **(c)**.

**Table 5 Magnetic, induced electric, and induced mechanical parameters at the side surface of a cylindrical culture well according to the Faraday-Neumann-Lenz and Lorentz laws (z = 4.5 cm)**

|  | t = 0.64 ms | Left neighborhood of t = 1.36 ms | Right neighborhood of t = 1.36 ms |
| --- | --- | --- | --- |
| $B_z$ [mT] | 2.0 | 3.1 | 3.1 |
| $B_x$ [mT] | Negligible | Negligible | Negligible |
| $B_y$ [mT] | Negligible | Negligible | Negligible |
| $|dB/dt|$ [T/s] | 2.2 | 1.0 | 2.6 |
| $|J|$, induced current density [mA/m$^2$] | 15.2 | 6.9 | 17.9 |
| $|F|$, induced force [pN] | 2.7 (maximum compression) | 1.9 (compression) | 4.9 (maximum traction) |

Table note: $|J| = \frac{1}{2}\sigma r |dB/dt|$, $|F| = \frac{1}{2}\pi h r^2 B_z |J|$ (physiological saline solution with electrical conductivity σ = 1.84 S/m and with height h = 1 mm; culture well radius r = 7.5 mm).

The preceding analytical solution was numerically confirmed and the forces were comparable to those applied in the study of cellular mechanics (Diz-Munoz et al., 2010), so, we could state that the seeded cells were also stimulated with time varying mechanical forces acting onto their plasma membrane at the frequency of 75 Hz. In addition, these forces belonged to planes parallel to the coils' planes and, consequently, under the tensegrity-mechanotransduction theory of Ingber (Mammoto and Ingber, 2010), they could be discomposed into their perpendicular/hydrostatic and tangent/shear components acting onto the cellular membranes.

## Discussion

It is well known that the physiological functions of cells and tissues can be influenced not only by molecules, but also by mechanical stimuli. In particular, according to the theory of Ingber (Ingber, 2003a; Ingber, 2003b; Ingber, 2006a; Ingber, 2006b; Mammoto and Ingber, 2010), during the *in vitro* culture inside bioreactors, the mechanical forces may change a specific cell status of force equilibrium, named isometric tensional prestress or "tensional integrity" or "tensegrity", inducing, via mechanotransduction, biochemical responses that may lead to changes to the transcriptional profile.

Inside our electromagnetic bioreactor, as shown above, the magnetic induction was able to elicit time varying mechanical forces acting perpendicularly or tangentially onto the cell membrane; as a consequence, these forces were able to modulate the cell tensegrity via tensile, compressive, and shear deformations.

Understanding how cells sense and react to mechanical forces has been shown to be crucial. For example, when osteoblasts are subjected to fluid shear stress, stretch-gated ion channels are opened and, due to the increased calcium concentration, numerous biochemical pathways are activated that lead to an enhanced transcription of bone matrix genes (Pavalko et al., 2003; Fassina et al., 2005; Young et al., 2009). In addition, both tension (i.e. traction) and compression affect the cell tensegrity: these forces alter the activities of intracellular signaling molecules such as Rho GTPases, guanine nucleotide exchange factors, GTPase activating proteins, and the MAPK pathway, consequently modulating the expression of transcription factors essential for the homeostasis of bone, cartilage and tooth tissues (Mammoto et al., 2012). Tension and compression may also influence the transcription activity more rapidly when their action is transmitted directly into the nucleus via the cytoskeleton linked to nuclear envelop proteins (Kim et al., 2012).

The biological effects inside our electromagnetic bioreactor could be also explained via the opening of voltage-gated $Ca^{2+}$ channels in the cell membrane. In particular, the electromagnetic stimulation can raise the net $Ca^{2+}$ flux in human cells and, according to Pavalko's diffusion-controlled/solid-state signaling model (Pavalko et al., 2003), the increase in the cytosolic $Ca^{2+}$ concentration is the starting point for numerous biochemical pathways.

In conclusion, in this study, we have performed a detailed numerical dosimetry inside our extremely-low-frequency electromagnetic bioreactor which has been successfully used in *in vitro* Biotechnology and Tissue Engineering researches (Fassina et al., 2006; Fassina et al., 2007; Fassina et al., 2008; Fassina et al., 2009; Fassina et al., 2010; Saino et al., 2011; Osera et al., 2011; Ceccarelli et al., 2013). The numerical dosimetry permitted to map the magnetic induction and to discuss its biological effects in terms of electromagnetically induced mechanical forces. In fact, the finite element method was shown to be effective in field calculations for a broad range of engineering (Di Barba et al., 2012) and of bioengineering (Di Barba et al., 2007; Di Barba et al., 2009; Di Barba et al., 2011) applications. So, in the intriguing frame of the tensegrity-mechanotransduction theory (Mammoto and Ingber, 2010), the study of these electromagnetically induced mechanical forces could be, in our opinion, a powerful tool to understand some effects of the electromagnetic stimulation whose mechanisms remain still elusive.

**Competing interests**
On behalf of all authors, the corresponding author states that there is no conflict of interests.

**Authors' contributions**
All authors contributed to the scientific design of the research, read and approved the final manuscript. LF and MEM wrote the manuscript. In addition, MEM performed the electric measurements, devised the finite element models, and did the numerical simulations. In addition, LF conceived the theoretical link between the electromagnetically induced mechanical forces and the biological mechanisms of the cell tensegrity.

**Acknowledgments**
The authors are grateful to Dr. R. Cadossi and Dr. S. Setti, who provided us, generously, the Biostim SPT Pulse Generator (Igea, Carpi, Italy). The authors are also grateful to Infolytica Corporation (Montréal, Canada), who provided us, generously, the MagNet software for the finite element analysis. The research was funded by the INAIL Grants 2010 to LF and FN.

**Author details**
[1]Dipartimento di Ingegneria Industriale e dell'Informazione, Università di Pavia, Via Ferrata 1, Pavia 27100, Italy. [2]Centro di Ingegneria Tissutale (C.I.T.), Università di Pavia, Pavia, Italy. [3]Dipartimento di Medicina Sperimentale, Università "Sapienza", Rome, Italy. [4]Dipartimento di Scienze Anatomiche, Istologiche, Medico-Legali e dell'Apparato Locomotore, Università "Sapienza", Rome, Italy.